\documentclass[conference]{IEEEtran}
\IEEEoverridecommandlockouts
\usepackage[numbers,sort&compress]{natbib}
\usepackage{amsmath,amssymb,amsfonts}
\usepackage{algorithmic}
\usepackage{graphicx}
\usepackage{textcomp}
\usepackage{xcolor}
\usepackage{multirow}
\usepackage{makecell}
\usepackage{hyperref}
\usepackage{threeparttable}
\def\BibTeX{{\rm B\kern-.05em{\sc i\kern-.025em b}\kern-.08em
    T\kern-.1667em\lower.7ex\hbox{E}\kern-.125emX}}
\begin{document}

\title{Single-channel EEG completion using Cascade Transformer\\
}

\author{\IEEEauthorblockN{
Chao Zhang\IEEEauthorrefmark{1}, 
Siqi Han\IEEEauthorrefmark{2}, 
Milin Zhang\IEEEauthorrefmark{1}\IEEEauthorrefmark{3}\IEEEauthorrefmark{4}}
\IEEEauthorblockA{\IEEEauthorrefmark{1}Department of Electronic Engineering, Tsinghua University,\\
\IEEEauthorrefmark{2}School of Modern Post (School of Automation), Beijing University of Posts and Telecommunications,\\
\IEEEauthorrefmark{3}Institute for Precision Medicine, Tsinghua University,\\
\IEEEauthorrefmark{4}Beijing National Research Center for Information Science and Technology, Tsinghua University\\
Corresponding author email: zhangmilin@tsinghua.edu.cn}}


\maketitle

\begin{abstract}
It is easy for the electroencephalogram (EEG) signal to be incomplete due to packet loss, electrode falling off, etc.
This paper proposed a Cascade Transformer architecture and a loss weighting method for the single-channel EEG completion, which reduced the Normalized Root Mean Square Error (NRMSE) by 2.8\% and 8.5\%, respectively. 
With the percentage of the missing points ranging from 1\% to 50\%, the proposed method achieved a NRMSE from 0.026 to 0.063, which aligned with the state-of-the-art multi-channel completion solution.
The proposed work shows it's feasible to perform the EEG completion with only single-channel EEG.
\end{abstract}


\section{Introduction}
A brain-machine interface (BMI) aims to establish a direct communication pathway between the nervous system of living creature and the external machine \cite{chaudhary2016brain, zhang2020electronic}.
Different types of neural signals have been demonstrated in various BMI applications.
Electroencephalogram (EEG) is one of the most popular electrophysiological signal in BMI scenarios such as attention evaluation \cite{Salenet22}, motor imagery \cite{tabar2016novel}, sleep staging \cite{MCMD22} and seizure detection \cite{tang2021design}. However, the EEG recording process is easily corrupted by the packet loss in wireless transmission, the
unexpected movement of subjects or the poor contact of electrodes, resulting in an incomplete signal.

Tensor completion methods (TCMs) have been proposed in several literature to perform the EEG completion by treating the recorded EEG as multi-channel tensor \cite{zhang2016removal,sole2018brain,duan2021robustness,akmal2021artificial}.
\cite{duan2021robustness} demonstrated that the simultaneous tensor decomposition and completion (STDC) could achieve a better and more robust performance among several TCMs.
The TCM family could discover a low-rank representation of the multi-channel signal, which could be further used in signal recovery. However, the TCM family relies on multiple EEG channels, which does not work for single-channel EEG recording.

Sequence to sequence neural network is another solution for EEG completion. \cite{el2019gate} introduced deep learning in this field with a gate-layer autoencoder (GLAE). The GLAE added a switch layer before the common autoencoder. The switch layer masked several input points during training. The model learned to complete the masked points according to the unmasked points. The GLAE achieved a RMSE level from 0.02 to 0.05 on two steady-state visually evoked potentials (SSVEP) EEG datasets. However, this model requires an eight-channel input. It does not work for single-channel EEG completion.

The Transformer \cite{vaswani2017attention} is a revolutionary sequence to sequence architecture which only consists of self-attention and feed forward network. Inspired by the success of Transformer in the fields such like natural language processing (NLP) \cite{devlin2018bert} and computer vision (CV) \cite{dosovitskiy2020image},
we propose to employ a Cascade Transformer architecture with a weighted loss for the single-channel EEG completion task in this work.
The contributions of this work are listed as follows:
\begin{itemize}
\item [a)]
This paper provides the first single-channel EEG completion solution in the literature. EEG completion is proven to be a feasible task with the absence of strong related adjacent channels.
\item [b)]
This paper proposes a Cascade Transformer architecture and a loss weighting method, which contribute 2.8\% and 8.5\% NRMSE decrease, respectively.
\item [c)]
The proposed method achieves a NRMSE from 0.026 to 0.063, which aligns with state-of-the-art multi-channel completion solution. In addition to the NRMSE, the frequency domain NRMSE (FD-NRMSE) for more fair cross dataset comparison is also reported.
\end{itemize}

The rest of the paper is organized as follows.
Section II introduces the proposed single-channel EEG completion method.
Section III illustrates the experimental results, while Section IV concludes the entire work.

\section{Single-channel EEG completion method}
\begin{figure*}[thbp]
    \centering
    \includegraphics[width=0.95\textwidth]{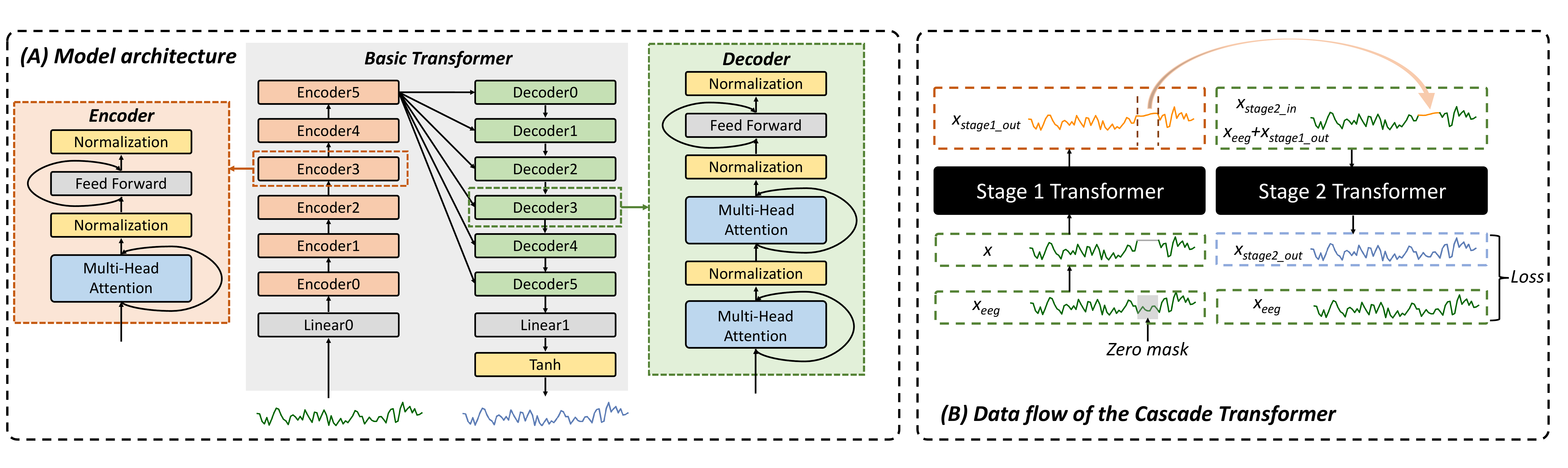} 
    \caption{The proposed Cascade Transformer.} 
    \label{fig:model}
\end{figure*}
\subsection{EEG completion task}
A single-channel incomplete EEG segment $x$ can be defined as
\begin{equation}
\label{eq:incompletion}
x[n]=\left\{
\begin{aligned}
x_{eeg}[n] & , & n \notin \{n_{missing}\} \\
x_{mask}[n] & , & n \in \{n_{missing}\}
\end{aligned}
\right.
\end{equation}
where $\{n_{missing}\}$ is the set of the indexes of all missing points. The symbol $x_{eeg}$ denotes the complete real EEG segment. The symbol $x_{mask}$ is the mask value of the missing points, which is usually set as a constant of 0.

Given an incomplete EEG segment, the single-channel EEG completion task in this work can be defined as
\begin{equation}
\label{eq:argmin}
\begin{aligned}
W_0 = \mathop{argmin}\limits_{{W}} (RMSE(x_{eeg}[n], x_{generate}[n])), \\ n \in \{n_{missing}\}
\end{aligned}
\end{equation}
where $x_{generate}$ denotes the generated signal of the completion model with weights $W$ and the optimal weights $W_0$.
\begin{equation}
\label{eq:RMSE}
\begin{aligned}
RMSE = \sqrt{\frac{\sum\limits_{n\in{n_{missing}}}(x_{eeg}[n]-x_{generated}[n])^2}{|\{n_{missing}\}|}}
\end{aligned}
\end{equation}
The RMSE shown in Eq. (\ref{eq:RMSE}) is recruited to evaluate the distance between the real EEG points and the generated EEG points, which can be replaced by other distance measures, such as L1-loss.

According to Eq. (\ref{eq:incompletion}) and (\ref{eq:argmin}), the EEG completion task is feasible to be handled with a sequence to sequence model. 
Define the input signal and the target output as $x$ and $x_{eeg}$, respectively. 
The model can learn to generate the masked EEG points based on the unmasked points in $x$ under a distance related loss function, such as the mean squared error (MSE) loss defined as
\begin{equation}
\label{eq:mseloss}
\begin{aligned}
loss_{mse}[n] = (output[n] - target[n])^2,  
\\ n = 0,...,N-1
\end{aligned}
\end{equation}
where $N$ is the length of the EEG segment $x$.

\subsection{Cascade Transformer}
The proposed Cascade Transformer is based on the Transformer in \cite{vaswani2017attention} as shown in Fig. \ref{fig:model}. In this part, it is necessary to briefly introduce several basic concepts of the Transformer before introducing the Cascade Transformer.

The self-attention as shown in Eq. (\ref{eq:attention}) is the key component of the Transformer.
\begin{equation}
\label{eq:attention}
\begin{aligned}
Attention(Q,K,V) = softmax(\frac{QK^T}{\sqrt{d_k}})V
\end{aligned}
\end{equation}
The $d_k$ is the length of $K$ for normalization.
The $Q$, $K$ and $V$ are the query, key and value vectors, respectively, which can be obtained as
\begin{equation}
\label{eq:QKV}
\left\{
\begin{aligned}
Q = x_{att}W_Q \\
K = x_{att}W_K \\
V = x_{att}W_V \\
\end{aligned}
\right.
\end{equation}
where $W_Q$, $W_K$ and $W_V$ are trainable weights of the model.
The symbol $x_{att}$ denotes the input tensor of the self-attention block. The multi-head attention layer in Fig. \ref{fig:model}(A) is an ensemble of $h$ self-attention blocks, where $h$ is the number of the heads.

As demonstrated in Fig \ref{fig:model}(A), a Transformer consists of several encoders and decoders. There are one multi-head attention block and one feed forward block in each encoder.
Each decoder consists of two multi-head attention blocks and one feed forward block. A feed forward block contains two linear layers. In order to enhance the robustness during training, several short-cuts are inserted in both the encoder and the decoder.
When it comes to data flow, each encoder processes the output of its former layer, while each decoder handles both the output of its former layer and the output of the last encoder.

In the proposed method, both the encoder number and the decoder number are set as 6. The sizes of the query, the key and the value vectors are all chosen as 16. 
Considering the intrinsic order of the EEG series, the position encoding layer in the original Transformer is removed, while the embedding layer is replaced by a linear layer.
In order to generate lifelike EEG with negative values, the Softmax layer in the original Transformer is replaced by a Tanh layer.

Fig. \ref{fig:model}(B) illustrates the proposed Cascade Transformer architecture which consists of two basic Transformer models.
When performing the EEG completion task, the missing points are masked as 0.
The entire EEG segment is fed into the Stage 1 Transformer. The Stage 1 Transformer generates a coarse estimation $x_{stage1\_out}$. Since the non-missing part of $x$ is always available, the input of Stage 2 Transformer can be formed as
\begin{equation}
\label{eq:stage2}
x_{stage2\_in}[n]=\left\{
\begin{aligned}
x_{eeg}[n] & , & n \notin \{n_{missing}\} \\
x_{stage1\_out}[n] & , & n \in \{n_{missing}\}
\end{aligned}
\right.
\end{equation}
Compared to $x$ in Eq. (\ref{eq:incompletion}), the input of the Stage 1 Transformer, the $x_{stage2\_in}$ is expected to achieve a better completion since the values on the missing indexes are more similar with the target EEG signal.

\subsection{Loss weighting}
According to Eq. (\ref{eq:argmin}), the RMSE calculated for the missing points between the real EEG and the generated signal is important to evaluate the performance. However, a common MSE loss in Eq. (\ref{eq:mseloss}) treats all output points equally. There are two aspects to consider while designing a loss function. On one hand, the model should focus on generating vivid EEG on the masked indexes. On the other hand, although the generated points on the unmasked indexes would not be used, the generating quality of these points is still important since they provide the necessary boundary information needed for the missing points.
In order to deal with this dilemma, we propose a  re-weighted MSE loss function as
\begin{equation}
\label{eq:loss-weight}
loss_{weighted}[n] = \\
\left\{
\begin{aligned}
loss_{mse}[n], n \notin \{n_{missing}\} \\
\alpha loss_{mse}[n], n \in \{n_{missing}\}
\end{aligned}
\right.
\end{equation}
where $\alpha$ is the weighting factor. 
When $\alpha$ is set as 1, the weighted loss is equivalent to the MSE loss. When the $\alpha$ increases, the model focuses more on the missing points.
\subsection{Metrics}
This paper utilizes the NRMSE defined in \cite{duan2021robustness}, where the RMSE is normalized by the range of the EEG channel as
\begin{equation}
\label{eq:NRMSE}
\begin{aligned}
NRMSE = \frac{RMSE}{max(x_{eeg})-min(x_{eeg})}
\end{aligned}
\end{equation}
In order to evaluate the smoothness of the generated signal, this paper also reports a frequency domain NRMSE (FD-NRMSE) defined as
\begin{equation}
\label{eq:FD-NRMSE}
\begin{aligned}
FDNRMSE = \sqrt{\frac{\sum\limits_{k=0}^{K-1}(F_{eeg}[k]-F_{generated}[k])^2}{K}}
\end{aligned}
\end{equation}
where $F$ is the spectrum of the normalized EEG series. $K$ is the total points of the spectrum with a typical value of 50 when the length of the EEG segment is 100, which results in a 1Hz spectrum resolution. When a higher spectrum resolution is expected, a DFT/FFT with more points could be recruited.

\section{Experimental results}
\subsection{Dataset}
The first nights of the first 10 participants in the Sleep-EDF-v2013 Cassette \cite{kemp2000analysis,goldberger2000physiobank} were recruited as the dataset to evaluate the proposed method. The Sleep-EDF contains at least one whole night EEG for each participant, which provides enough data to train the models such as Transformer. 
Aligned with the state-of-the-art single-channel sleep staging works \cite{phan2020towards,MCMD22}, the Fpz-Cz channel sampled at 100Hz was used in this paper.
\subsection{The influences of the the missing positions and the mask methods}
\label{ss:pos-mask}
There are several factors that matter in EEG completion, such as the number of the missing points, the position of the missing points and the mask method for the missing points.
We fixed the EEG segment length at 1 second (i.e. 100 points) and investigated the number of missing points for 1, 5, 10, 20 and 50.
The positions of the missing points were set as the beginning, the middle and the ending of the 1s EEG segment, respectively.
We tested three different masks, namely zero mask, random mask and EEG mask. 
The random mask set the missing points as an uniform distribution within the EEG signal range, while the EEG mask filled the missing points with their adjacent segment.
\begin{figure}[thbp] 
    \centering
    \includegraphics[width=0.48\textwidth]{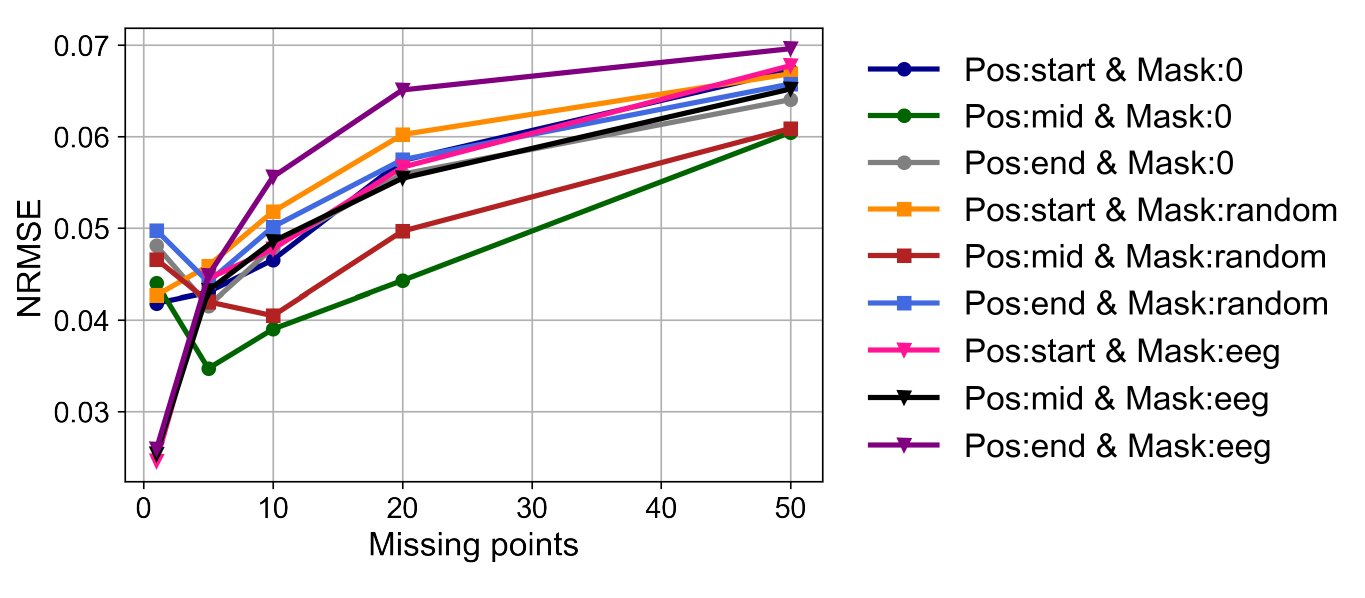} 
    \caption{Comparison of the NRMSEs under different missing percentages, positions and mask methods.} 
    \label{fig:pos-mask}
\end{figure}
The NRMSEs of the mentioned different settings are shown in Fig. \ref{fig:pos-mask}.
It could be seen that when the missing points number increases, the completion task becomes harder, resulting in a worse NRMSE.
It is better to use both the short segments before and after the missing points than to use a long segment only on one side.
As for the effect of different masks, Fig. \ref{fig:pos-mask} indicates that zero mask is enough for the model to generate a lifelike EEG. 

\subsection{Cascade Transformer}
\label{ss:cascade}
For the results in Section \ref{ss:pos-mask}, the missing position was set as middle with zero mask to evaluate the proposed cascade architecture.
\begin{figure}[thbp]
    \centering
    \includegraphics[width=0.48\textwidth]{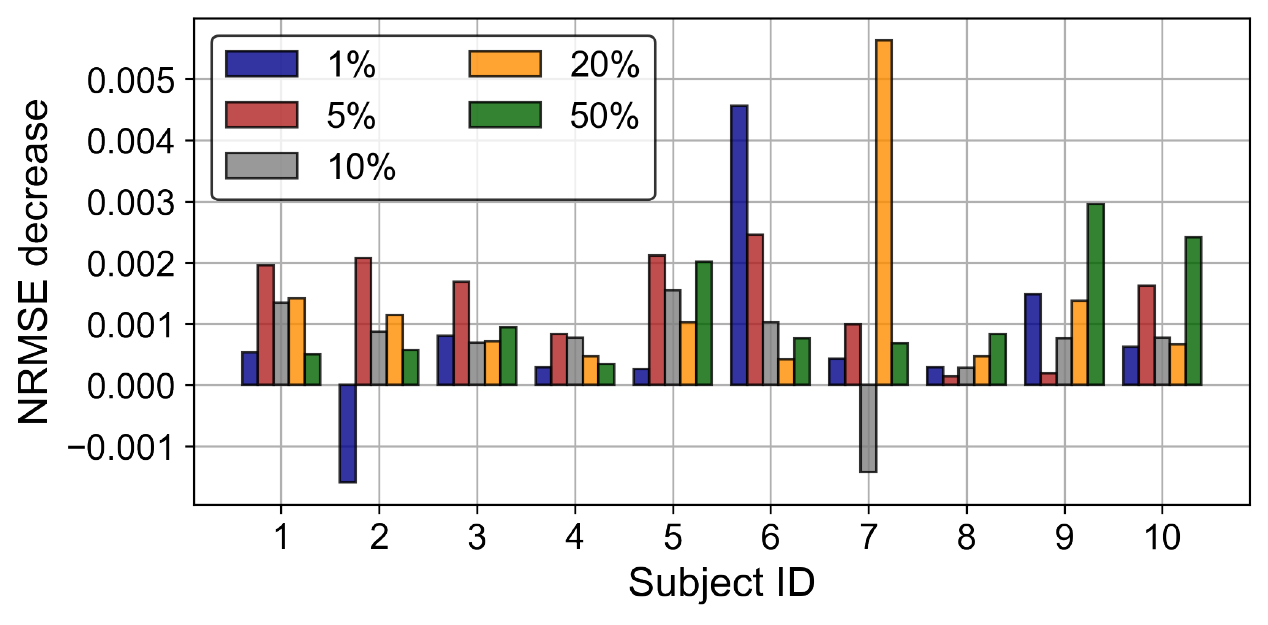} 
    \caption{The differences of the NRMSEs between the basic Transformer and the Cascade Transformer.}
    \label{fig:cascade_rmse}
\end{figure}
\begin{figure}[thbp]
    \centering
    \includegraphics[width=0.48\textwidth]{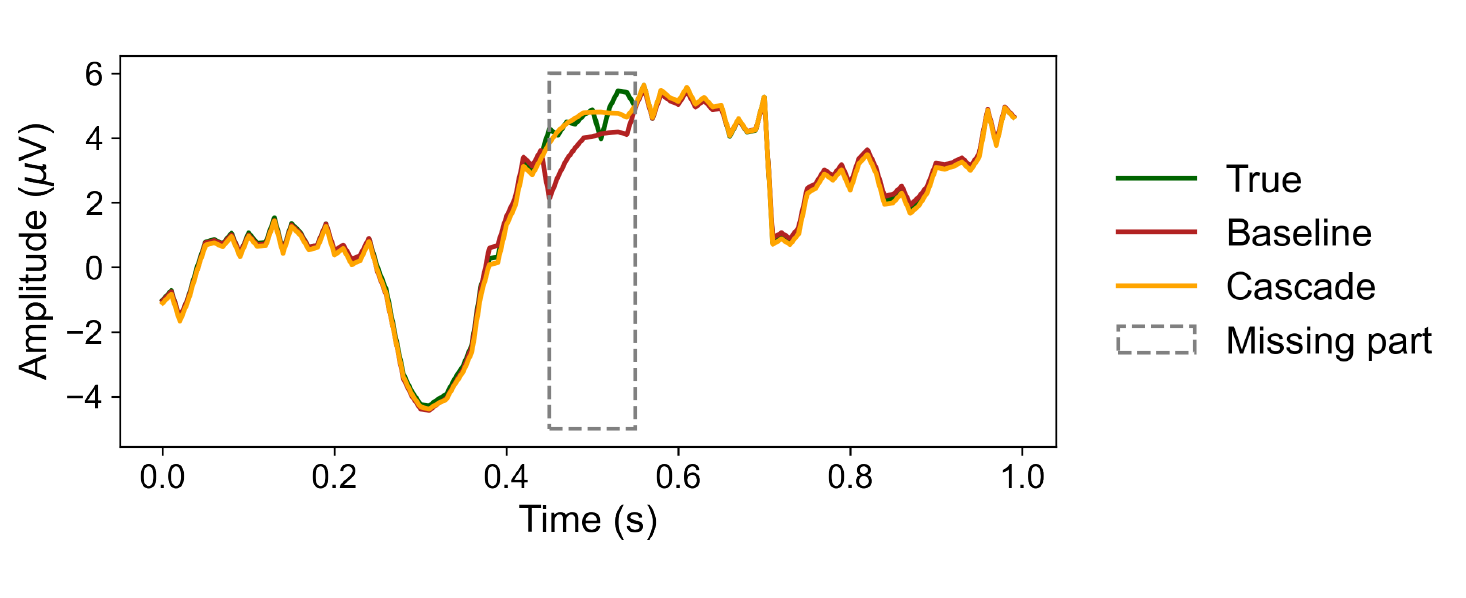} 
    \caption{An example of the completion results of the basic Transformer and the Cascade Transformer.}
    \label{fig:cascade_example}
\end{figure}
The differences of the NRMSEs between the basic Transformer and the Cascade Transformer under different percentages of the missing points are as shown in Fig. \ref{fig:cascade_rmse}.
It can be found that the cascade operation could benefit 48 of 50 trials with an average NRMSE decrease of 2.8\%.
Fig. \ref{fig:cascade_example} shows an example of the completion results. Compared to the results of the basic Transformer, the signal generated by the cascade architecture is significantly closer to the real EEG.

\subsection{Loss weighting}
\label{ss:loss-weight}
\begin{figure}[thbp]
    \centering
    \includegraphics[width=0.48\textwidth]{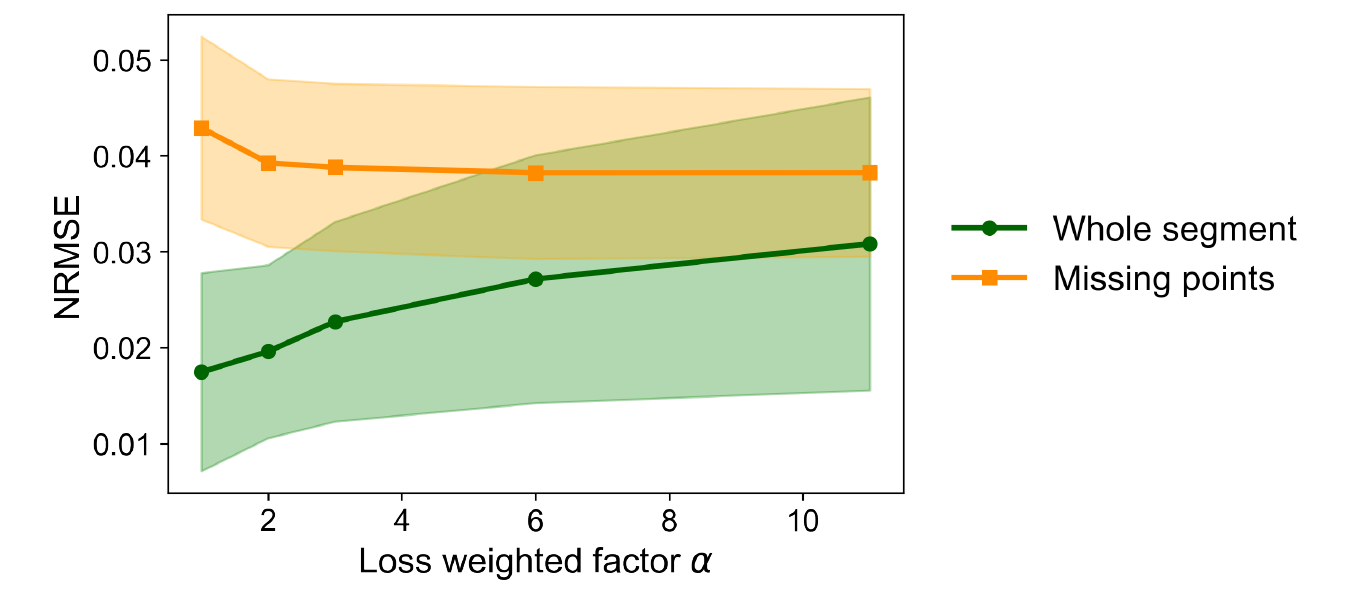} 
    \caption{The results under different loss weighting factors.}
    \label{fig:loss_weight}
\end{figure}

In order to investigate the influence on completion results of the loss weighting, five subjects were randomly selected with the weighting factor $\alpha$ referred to Eq. (\ref{eq:loss-weight}) set from 1 to 11. The percentage of the missing points was chosen as 5 and 10.
The average NRMSEs of the entire EEG segment and the missing part under different weighting factors were shown in Fig. \ref{fig:loss_weight}.
As the weighting factor $\alpha$ increases, the average NRMSE of the entire EEG segment increases, while the average NRMSE on the missing indexes decreases since more attention is focused on the missing points.
The average NRMSE on the missing indexes drops around 8.5\% when the value of $\alpha$ increases from 1 to 2. When the $\alpha$ is larger than 2, the average NRMSE tends to be smoother and the maximum decrease percentage is only 10.9\%.
Considering the risk of under-fitting, the loss weighting factor $\alpha$ is chosen as 2 in the further experiments.

\subsection{Comparison with the state-of-the-art}
The RMSE, NRMSE and FD-NRMSE with different numbers of missing points are as shown in Table \ref{tb:metrics}.
\begin{table}[!ht]
\caption{The metrics of the proposed method}
\label{tb:metrics}
\begin{center}
\scriptsize
\begin{threeparttable}
\scalebox{1.0}{
\begin{tabular}{c|ccc}
\hline
\multirow{2}{*}{Missing points} & \multicolumn{3}{c}{Metrics} \\
& RMSE & NRMSE & FD-NRMSE \\
\hline
1 & 9.46e-6$\pm$4.53e-6 & 0.026$\pm$0.012 & 0.076$\pm$0.042 \\
\hline
5 & 1.18e-5$\pm$4.65e-6 & 0.032$\pm$0.012 & 0.123$\pm$0.052 \\
\hline
10 & 1.38e-5$\pm$5.42e-6 & 0.038$\pm$0.014 & 0.172$\pm$0.073 \\
\hline
20 & 1.72e-5$\pm$6.00e-6 & 0.047$\pm$0.015 & 0.254$\pm$0.099 \\
\hline
50 & 2.30e-5$\pm$7.36e-6 & 0.063$\pm$0.019 & 0.389$\pm$0.141 \\
\hline
\end{tabular}}
\end{threeparttable}
\end{center}
\end{table}
Table \ref{tb:SOTA} compares the results between the proposed method and state-of-the-art multi-channel completion methods.
\begin{table}[!ht]
\caption{Comparison with the state-of-the-art EEG completion methods}
\label{tb:SOTA}
\begin{center}
\scriptsize
\begin{threeparttable}
\scalebox{1.0}{
\begin{tabular}{|c|c|c|c|}
\hline
& El-Fiqi \cite{el2019gate} & Duan \cite{duan2021robustness} & \textbf{This work} \\
\hline
Publication & IJCNN & SCTC & \textbf{BioCAS} \\
\hline
Year & 2019 & 2021 & \textbf{2022} \\
\hline
Dataset & SSVEP-EEG & MD-EEG\tnote{a} & \textbf{Sleep-EDF} \\
\hline
Sample rate & 128Hz & 256Hz & \textbf{100Hz} \\
\hline
EEG duration & 1s & 3s & \textbf{1s} \\
\hline
\#Channel & 8 & 11 & \textbf{1} \\
\hline
\#Subject & 10 & 15 & \textbf{10} \\
\hline
Method & \makecell[c]{Gate-layer\\ Autoencoder} & STDC & \textbf{\makecell[c]{Cascade\\ Transformer}} \\
\hline
Missing percentage & $<$25\% & $<$3.03\% & \textbf{$<$50\%} \\
\hline
RMSE & 0.022$\sim$0.037 & - & \textbf{9.46e-6$\sim$2.30e-5} \\
\hline
NRMSE & 0.068$\sim$0.115\tnote{b} & 0.006$\sim$0.016\tnote{c} & \textbf{0.026$\sim$0.063} \\
\hline
FD-NRMSE\tnote & - & - & \textbf{0.076$\sim$0.389} \\
\hline
\end{tabular}}
 \begin{tablenotes}
        \footnotesize
        \item[a] Motor decoding EEG.
        \item[b] Estimated by the open-source dataset \cite{kalunga2016online} in \cite{el2019gate}.
        \item[c] Calculated by the Log-NRMSE (1.8$\sim$2.2) in \cite{duan2021robustness}.
      \end{tablenotes}
    \end{threeparttable}
\end{center}
\end{table}
The proposed method achieved a NRMSE from 0.026 to 0.063 when the percentage of missing points ranged from 1\% to 50\% utilizing only single-channel EEG, which outperformed the autoencoder model \cite{el2019gate}. 
The STDC \cite{duan2021robustness} achieved the lowest NRMSE with a less than 3.03\% percentage of missing points. However, there was no evaluation results when more points were missed in one segment.
In addition, the proposed method showed a FD-NRMSE from 0.076 to 0.389, which could be taken into comparison in future research.

\section{Conclusion}
This paper proposed a Cascade Transformer architecture and a loss weighting method for single-channel EEG completion, which decreased the NRMSE by 2.8\% and 8.5\%, respectively.
The proposed architecture featured a NRMSE from 0.026 to 0.063 with the percentage of the missing points ranging from 1\% to 50\%, which aligned with the state-of-the-art multi-channel EEG completion methods. In order to make a more fair and diverse cross dataset comparison, this paper proposed to also report the FD-NRMSE.
As far as the authors know, this paper is the first work for single-channel EEG completion in the literature.

\section*{Acknowledgment}
This work is supported in part by the National Key Research and Development Program of China (No.2018YFB220200*), in part by the Natural Science Foundation of China through grant 92164202. in part by the Beijing Innovation Center for Future Chip.

\small
\bibliographystyle{IEEEtran}
\bibliography{completion}


\end{document}